\documentclass[
10pt, % Main document font size
a4paper, % Paper type, use 'letterpaper' for US Letter paper
oneside, % One page layout (no page indentation)
headinclude,footinclude, % Extra spacing for the header and footer
BCOR5mm, % Binding correction
]{scrartcl}

\usepackage[
nochapters, % Turn off chapters since this is an article        
beramono, % Use the Bera Mono font for monospaced text (\texttt)
eulermath,% Use the Euler font for mathematics
pdfspacing, % Makes use of pdftex’ letter spacing capabilities via the microtype package
dottedtoc % Dotted lines leading to the page numbers in the table of contents
]{classicthesis} % The layout is based on the Classic Thesis style

\usepackage{arsclassica} % Modifies the Classic Thesis package

\usepackage[T1]{fontenc} % Use 8-bit encoding that has 256 glyphs

\usepackage[utf8]{inputenc} % Required for including letters with accents

\usepackage{graphicx} % Required for including images
\graphicspath{{Figures/}} % Set the default folder for images

\usepackage{enumitem} % Required for manipulating the whitespace between and within lists

\usepackage{lipsum} % Used for inserting dummy 'Lorem ipsum' text into the template

\usepackage{subfig} % Required for creating figures with multiple parts (subfigures)

\usepackage{amsmath,amssymb,amsthm} % For including math equations, theorems, symbols, etc

\usepackage{varioref} % More descriptive referencing

\theoremstyle{definition} % Define theorem styles here based on the definition style (used for definitions and examples)

\theoremstyle{plain} % Define theorem styles here based on the plain style (used for theorems, lemmas, propositions)

\theoremstyle{remark} % Define theorem styles here based on the remark style (used for remarks and notes)

\hypersetup{
colorlinks=true, breaklinks=true, bookmarks=true,bookmarksnumbered,
urlcolor=webbrown, linkcolor=RoyalBlue, citecolor=webgreen, % Link colors
pdftitle={}, % PDF title
pdfauthor={\textcopyright}, % PDF Author
pdfsubject={}, % PDF Subject
pdfkeywords={}, % PDF Keywords
pdfcreator={pdfLaTeX}, % PDF Creator
pdfproducer={LaTeX with hyperref and ClassicThesis} % PDF producer
} % Include the structure.tex file which specified the document structure and layout

\title{\normalfont\spacedallcaps{Torrent Driven (TD) Coin: A Crypto Coin with Built In Distributed Data Storage System}} % The article title

\author{\spacedlowsmallcaps{Anirudha Paul}} % The article author(s) - author affiliations need to be specified in the AUTHOR AFFILIATIONS block

\date{} % An optional date to appear under the author(s)

\begin{document}

%----------------------------------------------------------------------------------------
%	HEADERS
%----------------------------------------------------------------------------------------

\renewcommand{\sectionmark}[1]{\markright{\spacedlowsmallcaps{#1}}} % The header for all pages (oneside) or for even pages (twoside)
\lehead{\mbox{\llap{\small\thepage\kern1em\color{halfgray} \vline}\color{halfgray}\hspace{0.5em}\rightmark\hfil}} % The header style

\pagestyle{scrheadings} % Enable the headers specified in this block

%----------------------------------------------------------------------------------------
%	TABLE OF CONTENTS & LISTS OF FIGURES AND TABLES
%----------------------------------------------------------------------------------------

\maketitle % Print the title/author/date block

% \setcounter{tocdepth}{2} % Set the depth of the table of contents to show sections and subsections only

% \tableofcontents % Print the table of contents

% \listoffigures % Print the list of figures

% \listoftables % Print the list of tables

%----------------------------------------------------------------------------------------
%	ABSTRACT
%----------------------------------------------------------------------------------------

\section*{Abstract} % This section will not appear in the table of contents due to the star (\section*)

In recent years decentralized currencies developed through Blockchains are increasingly becoming popular because of their transparent nature and absence of a central controlling authority. Though a lot of computation power, disk space, and energy are being used to run this system, most of these resources are dedicated to just keeping the bad actors away by using Proof of Work, Proof of Stake, Proof of Space, etc., consensus. In this paper, we discuss a way to combine those consensus mechanism and modify the defense system to create actual values for the end-users by providing a solution for securely storing their data in a decentralized manner without compromising the integrity of the blockchain. 

%----------------------------------------------------------------------------------------
%	AUTHOR AFFILIATIONS
%----------------------------------------------------------------------------------------

% \let\thefootnote\relax\footnotetext{* \textit{Department of Biology, University of Examples, London, United Kingdom}}

% \let\thefootnote\relax\footnotetext{\textsuperscript{1} \textit{Department of Chemistry, University of Examples, London, United Kingdom}}

%----------------------------------------------------------------------------------------

% \newpage % Start the article content on the second page, remove this if you have a longer abstract that goes onto the second page

%----------------------------------------------------------------------------------------
%	INTRODUCTION
%----------------------------------------------------------------------------------------

\section{Introduction}

When Bitcoin \cite{nakamoto2012bitcoin} was first introduced, its Proof of Work consensus showed us a different way to combat adversaries. Rather than blocking the adversaries directly, it made an attack on the network extremely expensive. Though this system is preventing any major attack on the system, the tendency of centralized mining is increasing exponentially. Nowadays, ordinary people can not participate in bitcoin mining. As a result, the system is not as democratized as initially thought. 

Another widespread consensus is Proof of Stake, which is adapted in Ethereum 2.0 \cite{Buterin2013}, Cardano \cite{kiayias2017ouroboros}, etc., coins, where you stake your own money to participate. Any bad behavior can cost the miners loosing their staked money. Though it is power efficient, staking actual currency without doing any work can cause nothing at stake, whale problems, etc. 

To address this, we propose a consensus that uses this distributed network of the miner to store user data. By securely storing this data and continuously proving the proof of storage, the miners will earn seed points. Instead of staking the currency they maintain, they will use these earned seed points to claim a spot in the mining round. In a nutshell, miners should provide some utility to earn points that enable them to mine further to get actual cryptocurrency. Spoofing the process of earning these seed points is hard enough so that honest miners have no incentive to deviate from the intended flow. And as the underlying mining technology is basically already tested proof of stake, the integrity of the network is as good as other similar currencies like Ethereum 2.0, Cardano, etc.

\section{Previous Whitepaper Shortcomings}

In the Bitcoin whitepaper \cite{nakamoto2012bitcoin}, the author introduced the idea of Proof Of Work, where miners need to continuously try different nonce to generate a hash with the required number of zero bit. But as time goes by, this challenge of generating the intended hash has become so complex that nowadays it is near impossible to mine effectively without designated ASICS, which is a huge waste of money and resources for doing calculations that have no other purpose other than showing commitment. 

To address this, Ethereum is introducing Proof of Stake in its system. But it also introduces new problems - as now the "work" part is removed to show commitment, miners have the opportunity to sign multiple blocks from parallel chains, making it hard for forks to converge. Though Ethereum punishes this behavior by slashing the coin of bad actors, it is still not completely secured from manipulation by the whales who have the majority coin in control. 

Another consensus mechanism is proof of space \cite{cryptoeprint:2013:796}, where the miner fills up its disk space with garbage information generated by mathematical hash functions, and the verifier sends them challenges from time to time to validate whether the miner is holding the data or not. The issue is the disk space the data is taking is not meaningful. It is there to show commitment, just like Bitcoin's proof of work. 

We shall see later in this paper that it is possible to address some of the shortcomings mentioned above. 
%----------------------------------------------------------------------------------------
%	METHODS
%----------------------------------------------------------------------------------------

\section{Architecture}

The whole mechanism of this blockchain can be described with three sections. 

\begin{enumerate}[noitemsep] % [noitemsep] removes whitespace between the items for a compact look
\item Block structure
\item Consensus
\item Method of issuing the token
\end{enumerate}

The block structure follows the standard Bitcoin block structure. The consensus follows the proof of stake mechanism, but it is different in a sense it doesn't use the main coin as a stake but instead uses a secondary Seed Bonus Token. The main deviation comes in the method of issuing the tokens. Though the main Torrent Driven Coin (TD Coin) is issued with a standard block reward, the secondary coin (Seed Bonus Token) needed for staking and mining has a different structure for minting.

%------------------------------------------------

\subsection{Block structure}
\begin{figure}[tb]
    \centering 
    \includegraphics[width=0.6\columnwidth]{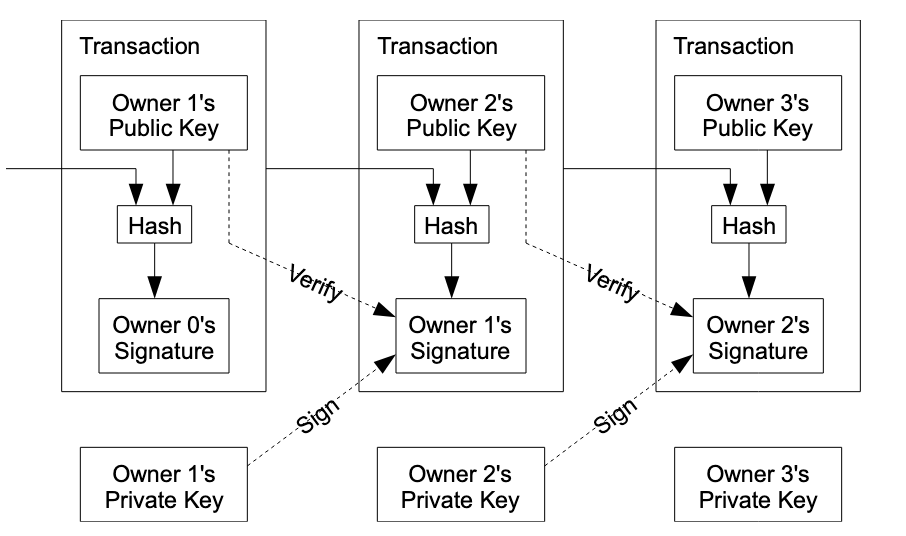} 
    \caption[Transaction structure]{Transaction structure (Standard Bitcoin) \cite{nakamoto2012bitcoin}} % The text in the square bracket is the caption for the list of figures while the text in the curly brackets is the figure caption
    \label{fig:trx} 
    \end{figure}

Transactions are arranged in standard Bitcoin format, where every transaction is spent from the coin, not from the account. The coin owner transfers the coin by digitally signing the hash of the previous transaction and adding the next owner's public key at the end. The new owner can verify the signatures by following the chain of ownership. And to avoid double-spending, only the oldest transaction of any coin is considered valid. \ref{fig:trx} 

\begin{figure}[tb]
    \centering 
    \includegraphics[width=0.6\columnwidth]{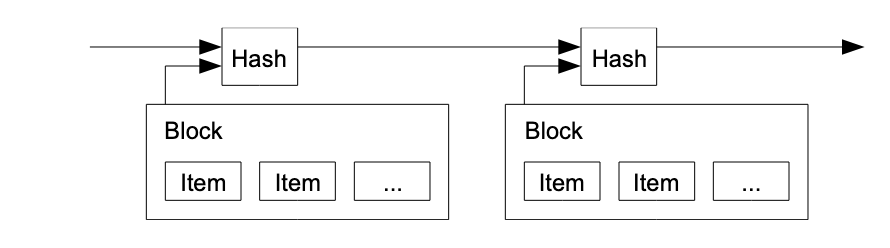} 
    \caption[Block structure]{Block structure (Standard Bitcoin) \cite{nakamoto2012bitcoin}} % The text in the square bracket is the caption for the list of figures while the text in the curly brackets is the figure caption
    \label{fig:block} 
    \end{figure}

In the chain, many of these transactions are arranged in a block, and their hash value links them. The size of the blocks can be dynamically adjusted like Ethereum based on the network congestion. \ref{fig:block}

\subsection{Consensus}

\begin{figure}[tb]
    \centering 
    \includegraphics[width=0.8\columnwidth]{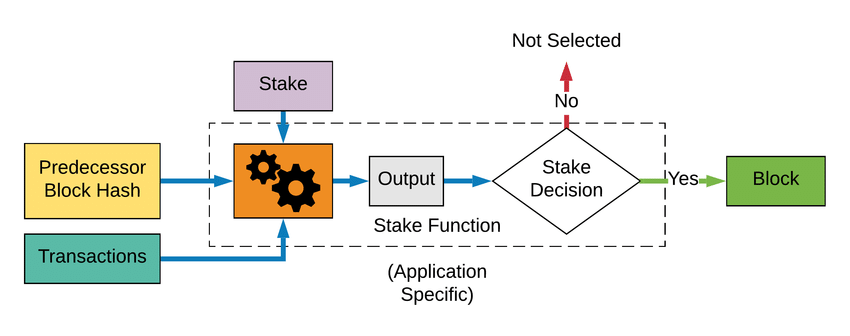} 
    \caption[A high level architecture of Proof of Stake]{A high level architecture of Proof of Stake consensus \cite{pospic}} % The text in the square bracket is the caption for the list of figures while the text in the curly brackets is the figure caption
    \label{fig:pos} 
    \end{figure}

In this modified proof of stake consensus mechanism \ref{fig:pos}, miners agree to lock up the whole amount of their secondary coin, "Seed bonus token," for getting the chance to validate new blocks of data to be added to a blockchain. 

The blockchain algorithm selects validators from the pool of queued miners based on how much seed bonus token their accounts have. The more seed bonus token a miner has, the better chance of being chosen to mine and earn newly minted primary crypto - "Torrent Driven Coin" as a reward if the block gets added to the main chain. A portion of their seed bonus token is burnt to encourage future data seeding, and the rest is returned back to their wallet again.

If a validator is caught cheating, they could be punished by burning all their seed bonus tokens and sending them to an unusable wallet address to which nobody has access, making them useless forever.

\subsection{Method of issuing the token}

There are three types of token present in this system.

% \lipsum[6] % Dummy text

\paragraph{1. Torrent Driven Coin} This is the standard-issue coin that can be exchanged between any parties present in the blockchain. It can only be minted by mining a block in this blockchain. Other than exchanging it as money, users can also burn an amount of this coin through a smart contract present in layer 1 of this blockchain to get the Leecher Token described below. The exchange rate between Torrent Driven Coin and Leecher Token will be adjusted dynamically based on supply and demand.

% \paragraph{Different Paragraph Description} \lipsum[8] % Dummy text

\paragraph{2. Leecher Token} Leecher token grants the ability to upload your data to other users or the ability to host other's data in your machine and get a seed bonus. The only way to get this token is to send an amount of Torrent Driven coin to the predefined smart contract, and the smart contract will burn the coin and give the sender an amount of Leecher Token in exchange. The exchange rate is controlled algorithmically to address the supply-demand issue. 

Each leech token grants access to upload or host one MB of data. If a user wants to host 30MB of data in one additional copy in the network, the process will work like this - 

For example, Alice wants to make a copy of her data on the peer-to-peer network, and Bob wants to host data for the seed bonus token.

\begin{figure}[tb]
    \centering 
    \includegraphics[width=0.8\columnwidth]{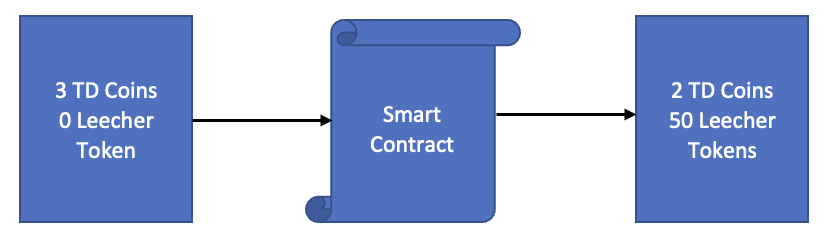} 
    \caption[Mint]{Minting Leech Token} % The text in the square bracket is the caption for the list of figures while the text in the curly brackets is the figure caption
    \label{fig:leech} 
    \end{figure}

First, Alice will send an amount of Torrent Driven coin to a specific smart contract to get 30 or more leecher tokens as seen in figure \ref{fig:leech}. 

\begin{figure}[tb]
    \centering 
    \includegraphics[width=0.8\columnwidth]{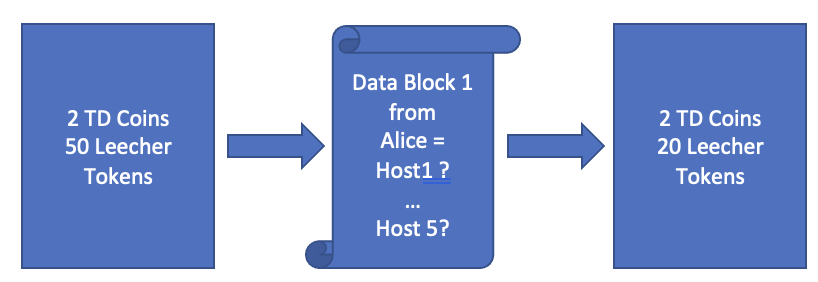} 
    \caption[Host]{Hosting Request} % The text in the square bracket is the caption for the list of figures while the text in the curly brackets is the figure caption
    \label{fig:hostingreq} 
    \end{figure}

Then she will join the pool of hosting requests to find possible hosts with the 30 leecher tokens in place like in figure \ref{fig:hostingreq}. 

On the other side, Bob also needs to exchange Torrent Driven coins to show commitment as a seeder. So, for example, if he has 50 leecher tokens, he can request the smart contract to match him with 50 * 5 = 250 data blocks. Five is a constant here, representing each data block will be saved by at least five seeders. 

\begin{figure}[tb]
    \centering 
    \includegraphics[width=0.8\columnwidth]{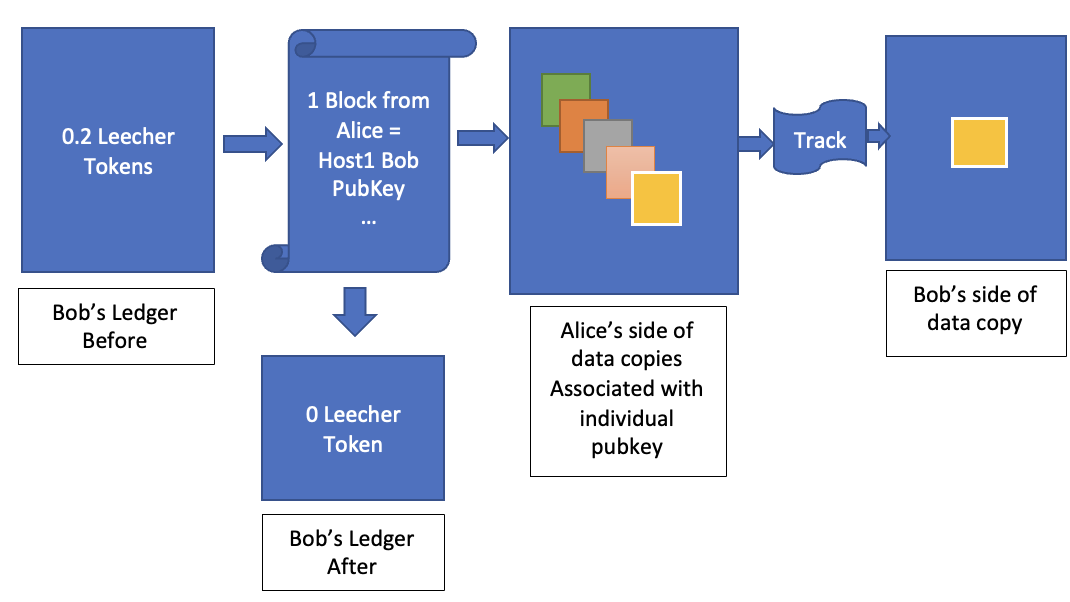} 
    \caption[Dist]{Data Distribution Process} % The text in the square bracket is the caption for the list of figures while the text in the curly brackets is the figure caption
    \label{fig:datadist} 
    \end{figure}

Then what the smart contract will do is match each block Alice wants to host with five different Host addresses. Alice will see all the public keys associated with each of her data blocks on the blockchain. The application on her side will make five copies of data, add the public key to those blocks, make a torrent tracker and publish the tracker on the blockchain. When the tracker gets published, the group of hosts, including Bob, can see the tracker and use that to only download the data attached with their individual public keys in header like in figure \ref{fig:datadist}.

\begin{figure}[tb]
    \centering 
    \includegraphics[width=0.3\columnwidth]{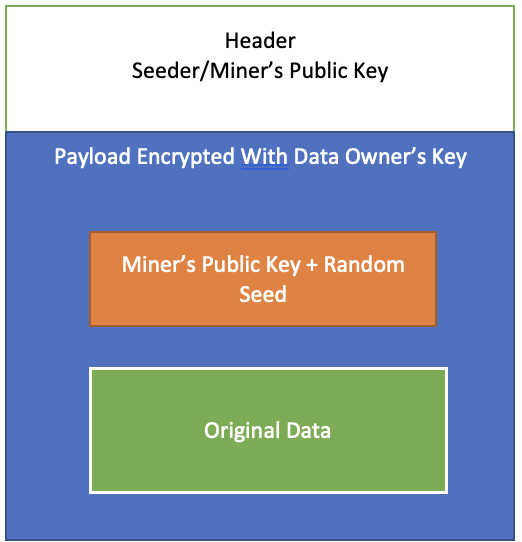} 
    \caption[payload]{Payload Structure} % The text in the square bracket is the caption for the list of figures while the text in the curly brackets is the figure caption
    \label{fig:payload} 
    \end{figure}

Not only that, each payload is also appended with the seeder's public key and random value before encryption which prevents seeders from swapping the payload between them. Each seeder has his own unique version of the same payload that the data owner himself can only decrypt. Sample payload structure can be seen in figure \ref{fig:payload}

The entire process will be covered with Byzantine Fault tolerance and other cross-chain swap techniques if any involved party deviates from the intended path like - going offline, not signing, false signing, etc. The smart contract will ensure that other parties won't have to consume the loss. 

So this leecher token is not exchangeable between addresses. It can only be used for hosting purposes.

%------------------------------------------------

\paragraph{3. Seed Bonus Token}

The sole purpose of this token is to ensure data integrity and facilitate staking for proof of stake. 

The first issue is - what is the guarantee the host is storing the data? The standard checksum used in torrent systems is not enough to solve this issue. Because to do that, they need to send the data back to the validator, which is slow and unnecessary. Instead, the original data owner or some other validator can save some part of that data in their system, and after random intervals, they can request zero-knowledge proof from the host. The time to response window is short to avoid data recreation. If they can't respond with the proof within that window, that proves the miner is either offline or is not actually storing the data - both of these behaviors are unacceptable. 

But if they can respond to the challenge properly along with able to send a small amount of data to show seed liveliness and passing the checksum test, the host will be rewarded with a small amount of new token - the "Seed Bonus Token." The download and checksum check is similar to a normal torrenting system \cite{cohen2003incentives}. And the zero-knowledge proof is similar to this proof of space and time described in this paper \cite{moran2019simple}. The whole process can be done in side chain and only publish the result in main chain to claim reward.

Now there is a new token in the account for potential miners - which is really hard to spoof and also not transferable. That's why the reason to be a bad actor to earn this token is very slim. So the possibility of the host honestly seeding the data is much higher. 

But now the question is, what is the incentive for them to earn this reward? The answer is - they can use this reward to stake to get a chance to validate a block and get the block reward in TD coin, which is actually transferable and can be used as actual currency. 

From this point on, the consensus falls back on the proof of stake. The difference here is - instead of staking the main currency, the miners are staking their off-chain hosting work reward earned through proof of space.

%----------------------------------------------------------------------------------------
%	RESULTS AND DISCUSSION
%----------------------------------------------------------------------------------------

\section{Scalability, Security, and Resource-Efficiency}

The system is scalable in a sense underlying everything; it is still using the proof of stake as its consensus, which is pretty scalable and secured under heavy transactions. One might think the whole process of downloading and uploading data can slow down the main blockchain - which is not true. Cause the upload, download, and zero-knowledge prove that part of the system can be considered off-chain work and side chain transactions. Only the result of each checkpoint is published on the main chain. So the main chain is not slowed down by all the bottlenecks associated with proof of space. 

The main network will continue to use proof of stake, but the staked coins are influenced by tokens earned in the independent torrenting mechanism and proof of space consensus. 

The whole system is resource-efficient because though miners have to do a lot more work, especially allocating more disk space than the traditional proof of stake, the silver lining is that those works are not wasted work. By using the storage, they are providing actual value to the users. There is no unnecessary computation power used like Proof of Work.

Overall the security is also up to par with other traditional coins. The original main network is using proof of stake, which after many iterations and research, is now pretty secure to be relied upon. The staked point is going through the zero-knowledge proof of space and byzantine consensus, making it extremely hard to profit from being a bad actor. There is no way to replicate the data as the packets are encrypted with individual miner's public key, and the only way to list your public key as a valid host is by burning actual Torrent Driven coins. As the exchange rate is controlled algorithmically, it is not easy to inflate or deflate the staking pool or make a hostile takeover.

So by combining all those techniques, the proposed blockchain is scalable, secured, and resource-efficient and provides actual value to users instead of the perceived value of traditional crypto coins. 

\section{Conclusion}
The primary goal of this white paper was to create intrinsic value for the whole distributed mining network of blockchain. By incorporating torrenting mechanisms in the proof of space framework, we have introduced a solution for securely and reliably storing multiple copies of data on the internet. Moreover, as there are monetary exchanges and interests involved in the process, the chance of abandoning the torrent by the seeder is really low. And it is also a great way to incentify hosting private encrypted data. In the process, it is now replacing the redundent generated data of Proof of Space with actual meaningful information.

By adding signatures, zero-knowledge proof, and byzantine consensus to track the seed status and facilitate a reward mechanism, we have minimized the risk of bad actors creating fake seed data and, in the process, attacking the staking pool.

On the other hand, removing staking coins with staking work has introduced a toned-down version of Proof of Work in the Proof of Stake mechanism, making the architecture secure like PoW and scalable like vanilla PoS. 

Finally, introducing a utility inside the network and regulating the exchange rate algorithmically has the possibility of reducing the current deflationary nature of the crypto coins where no one wants to use it in the real-world other than speculative investment. Because in this architecture, seeding more data is beneficial for the miners, that's why the competition among miners can reduce the cost for the storing data for the normal users.
%----------------------------------------------------------------------------------------
%	BIBLIOGRAPHY
%----------------------------------------------------------------------------------------
\newpage
\renewcommand{\refname}{\spacedlowsmallcaps{References}} % For modifying the bibliography heading

\bibliographystyle{unsrt}

\bibliography{sample.bib} % The file containing the bibliography

%----------------------------------------------------------------------------------------

\end{document}